\documentclass{ifacconf}

\usepackage{graphicx} % include this line if your document contains figures
\usepackage{epsfig}   % for postscript graphics files
\usepackage{amsmath}  % assumes amsmath package installed
\usepackage{amssymb}  % assumes amsmath package installed
\usepackage{natbib}   % required for bibliography
\usepackage{anyfontsize}
\usepackage[shortlabels]{enumitem}
\usepackage{calrsfs}
\usepackage{enumitem}
\usepackage{accents}
\usepackage{dsfont}
\usepackage{mathdots}
\usepackage{setspace}
\usepackage{color,soul}
\usepackage{algorithm}
\usepackage{algorithmicx}
\usepackage{algpseudocode}
\DeclareMathAlphabet{\pazocal}{OMS}{zplm}{m}{n}

\newcommand{\norm}[1]{\left\lVert#1\right\rVert}
\allowdisplaybreaks

\makeatletter
\def\endfigure{\end@float}
\def\endtable{\end@float}
\makeatother

\newcommand{\col}[1]{\text{col}\left(#1\right)}
\newcommand{\cov}[1]{\text{cov}\left(#1\right)}
\newcommand{\EE}[1]{\mathbb{E}\left[#1\right]}
\newcommand{\tr}[1]{\text{tr}\left(#1\right)}
\newcommand{\setdef}[2]{\left\{#1\,\left|\,#2\right.\right\}}

\let\ifacconfcaptionwidth\captionwidth
\usepackage[caption=false]{subfig}
\let\captionwidth\ifacconfcaptionwidth

\begin{document}
\begin{frontmatter}

\title{Stochastic Data-Driven Predictive Control: Regularization, Estimation, and Constraint Tightening} % \\

%\thanks[footnoteinfo]{This work was supported by the Swiss National Science Foundation under Grant 200021\_178890.}
%\thanks[copyright]{This work has been submitted to IFAC for possible publication.}
%\thanks[copyright]{\copyright\ 2020 the authors. This work has been accepted to IFAC for publication under a Creative Commons Licence CC-BY-NC-ND.}

\author[First]{Mingzhou Yin}
\author[Second]{Andrea Iannelli}
\author[First]{Roy S. Smith}

\address[First]{Automatic Control Laboratory, ETH Z\"{u}rich, Switzerland \\ (e-mail: \{myin,rsmith\}@control.ee.ethz.ch)}
\address[Second]{Institute for Systems Theory and Automatic Control, University of
Stuttgart, Germany (e-mail: andrea.iannelli@ist.uni-stuttgart.de)}

\begin{abstract}                % Abstract of not more than 250 words.
Data-driven predictive control methods based on the Willems' fundamental lemma have shown great success in recent years. These approaches use receding horizon predictive control with nonparametric data-driven predictors instead of model-based predictors. This study addresses three problems of applying such algorithms under unbounded stochastic uncertainties: 1) tuning-free regularizer design, 2) initial condition estimation, and 3) reliable constraint satisfaction, by using stochastic prediction error quantification. The regularizer is designed by leveraging the expected output cost. An initial condition estimator is proposed by filtering the measurements with the one-step-ahead stochastic data-driven prediction. A novel constraint-tightening method, using second-order cone constraints, is presented to ensure high-probability chance constraint satisfaction. Numerical results demonstrate that the proposed methods lead to satisfactory control performance in terms of both control cost and constraint satisfaction, with significantly improved initial condition estimation.
\end{abstract}

\begin{keyword}
data-driven control, model predictive control, estimation and filtering, stochastic optimal control, stochastic system identification
\end{keyword}
\end{frontmatter}

\section{Introduction}
Classical model-based control enables simple but powerful control design by considering typically parametric mathematical abstractions of system behaviors, known as models. However, this comes at the cost of additional modeling and identification effort, which constitutes the majority of the budget in model-based control design, in terms of both time and cost \citep{Hjalmarsson_2005}. In this regard, the concept of data-driven control provides an appealing alternative that designs the controller directly from data without parametric identification.

In this work, we focus on data-driven predictive control (DDPC), the data-driven counterpart to model predictive control (MPC). Similar to MPC, it solves a finite-horizon optimal control problem in a receding horizon fashion, but with nonparametric data-driven predictors instead of model-based predictors. Data-driven predictors for linear systems can be constructed by using the so-called Willems' fundamental lemma \citep{Willems_2005,IM-FD}, which characterizes all possible system behaviors with finite data.

While these predictors work well with deterministic data, showing equivalence to model-based design, they become ill-defined with stochastic data. Multiple works have stressed this issue by introducing an inner problem that finds the `optimal' predictor under some statistical principle, e.g., \cite{fiedler2021relationship,yin2020maximum,yin2021data,Breschi_2023}. This idea is known as indirect DDPC \citep{dorfler2022bridging}.

These stochastic data-driven predictors can be applied to DDPC design similar to the noise-free case. However, they suffer from the following problems with this certainty-equivalent implementation: 1) the control cost does not account for the prediction error, 2) the initial condition measurements suffer from noise which cannot be improved by collecting more data, and 3) the constraint satisfaction is not guaranteed. Aspects of these problems have been analyzed under the robust control framework where a bounded uncertainty set is prescribed \citep{Coulson_2019_reg,Berberich_2020,Huang_2023,Kl_ppelt_2022}. On the other hand, the situation is less clear under the stochastic control framework, where existing works adopt restrictive assumptions, such as noise-free offline data \citep{10171461} and exact polynomial chaos expansions of stochastic measurements \citep{pmlr-v211-pan23a}. This paper addresses these problems under general unbounded stochastic uncertainties, utilizing the prediction error quantification provided in \cite{yin2021data}.

In particular, three modifications are made to the certainty-equivalent DDPC algorithm. 1) The nominal control cost is replaced by the expected control cost. This introduces an additional uncertainty term that resembles the regularizer used empirically for DDPC problems, but here the weight is specified statistically without tuning. 2) The output initial condition is estimated by Kalman-filtering the output measurements with one-step-ahead predictions from the previous timestep. This significantly reduces the prediction error. 3) Output constraints are formulated as chance constraints and guaranteed by second-order cone (SOC) constraints. The effectiveness of these modifications is tested in a numerical example, where satisfactory control performance in terms of both control cost and constraint satisfaction is observed with significantly improved initial condition estimation.

\textit{Notation.} The expected value, standard deviation, and covariance are denoted by $\mathbb{E}[\cdot]$, $\text{std}(\cdot)$, and $\text{cov}(\cdot)$, respectively. The symbol $\text{Pr}(\cdot)$ indicates the probability of a random event. For a sequence of matrices $X_1,\dots,X_n$, we denote the row-wise and diagonal-wise concatenation by $\text{col}\left(X_1,\dots,X_n\right)$ and $\text{blkdiag}\left(X_1,\dots,X_n\right)$, respectively. $\text{diag}(\cdot)$ denotes the vector of diagonal elements of a square matrix. Given a signal $x:\mathbb{Z}\to \mathbb{R}^n$, its trajectory from $k$ to $k+N-1$ is indicated by $(x_i)_{i=k}^{k+N-1}=\text{col}(x_k,\dots,x_{k+N-1})$. For a vector $x$, $\norm{x}_P$ denotes the weighted $l_2$-norm $(x^\top Px)^{\frac{1}{2}}$. The rank and trace of a matrix are indicated by $\text{rank}(\cdot)$ and $\text{tr}(\cdot)$, respectively.

\section{Problem Formulation}
\label{sec:2}

Consider the observable part of a stable discrete-time linear time-invariant (LTI) dynamical system with disturbance and output noise, given by
\begin{equation}
\begin{cases}
x_{t+1}&=\ A x_t+B u_t+E w_t,\\
\hfil y_t&=\ C x_t + D u_t + v_t,
\label{eq:sys}
\end{cases}
\end{equation}
where $x_t \in \mathbb{R}^{n_x}$, $u_t \in \mathbb{R}^{n_u}$, $y_t \in \mathbb{R}^{n_y}$, $w_t \in \mathbb{R}^{n_w}$, $v_t \in \mathbb{R}^{n_y}$ are the states, inputs, outputs, disturbance, and output noise, respectively. The noise-free output is denoted by $y_t^0$. In this paper, the noise $v_t$ is considered to be zero-mean i.i.d. with covariance $\cov{v_t}=\sigma^2\mathbb{I}_{n_y}$.

The model parameters $(A,B,C,D,E)$ are unknown, but a matrix of input-disturbance-output trajectory data $Z=\begin{bmatrix}z_0^d&\cdots&z_{M-1}^d\end{bmatrix}$ has been collected, where each column
\begin{multline}
    z_i^d:=\text{col}\left(u_{t_i}^d,\dots,u_{t_i+L-1}^d,\right.\\\left. w_{t_i}^d,\dots,w_{t_i+L-1}^d,y_{t_i}^d,\dots,y_{t_i+L-1}^d\right)
    \label{eqn:zi}
\end{multline}
is a length-$L$ trajectory of the system, where the superscript $d$ denotes collected offline data. The availability of offline disturbance trajectories retroactively is commonly assumed in DDPC algorithms, e.g., \cite{pmlr-v211-pan23a}, and is practical in many applications. The noise in each column of outputs is assumed to be independent. This assumption holds exactly when the columns are separate trajectories or truncated from a longer trajectory with $t_{i+1}=t_i+L$, known as the Page construction \citep{iannelli2021design}. Another common construction of $Z$ is by choosing $t_{i+1}=t_i+1$, forming a block Hankel matrix. The matrix $Z$ is dubbed the signal matrix \citep{yin2020maximum}.

We are interested in designing a receding horizon control algorithm with a predictor derived from the signal matrix $Z$ instead of the model parameters. Such algorithms are known as indirect DDPC \citep{dorfler2022bridging}. Let $L=L_0+L'$, and $L_0$ be no smaller than the observability index of the system. In this paper, we consider the stochastic optimal tracking problem within a horizon of $L'$ that minimizes the following expected control cost
\begin{equation}
    \underset{(\hat{u}_k^t)_{k=0}^{L'-1}}{\text{min}}\ J_t:= \sum_{k=0}^{L'-1}\norm{\hat{u}_k^t}_R^2+ \EE{\sum_{k=0}^{L'-1}\norm{\hat{y}_k^t-r_{t+k}}_Q^2}
\end{equation}
at time $t$, where $\hat{u}_k^t$ is the designed input at time $(t+k)$, $\hat{y}_k^t$ is a random variable that predicts the noise-free future output $y_{t+k}^0$, $r_t$ denotes the reference trajectory, and $Q,R$ are the output and the input cost matrices respectively.

It is also desired to constrain the outputs within a polytopic set $\mathcal{Y}_t:=\setdef{y_t}{H^t y_t \leq q^t}$ at time $t$, where $H^t:= \left[h_1^t\ \dots\ h_{n_c}^t\right]^\top\in \mathbb{R}^{n_c\times n_y}$ and $q^t:=\col{q_1^t,\dots,q_{n_c}^t}\in\mathbb{R}^{n_c}$. However, due to the existence of unbounded noise, the output constraints can only be satisfied with high probability as chance constraints, which will be detailed in later sections. The input is constrained to be in the set $\mathcal{U}^t$ at time $t$, i.e.,
\begin{equation}
    \hat{u}_k^t\in \mathcal{U}_{t+k},\ \forall\,k=0,\dots,L'-1.
    \label{eqn:ucon}
\end{equation}
Then, the first element in the optimal input sequence is applied to the system, i.e., $u_t:=\hat{u}_0^t$ and the noisy output $y_t=y_t^0+v_t$ is measured.

There are multiple aspects to consider when solving this problem, which will be discussed in the following sections.
\begin{enumerate}[1)]
    \item There are uncertainties in both the prediction conditions and the signal matrix. An accurate predictor is desired under these uncertainties.
    \item The expected control cost needs to be formulated as a tractable objective function.
    \item Tractable formulations of the output constraints need to be derived.
\end{enumerate}

\section{Stochastic Data-Driven Predictor}

\subsection{Deterministic Prediction}

With sufficiently persistently exciting inputs, the range space of $Z$ contains all possible trajectories of the system in the noise-free case, i.e., $v_t=\mathbf{0}_{n_y}$. The following result derived from the Willems' fundamental lemma \citep{Willems_2005} provides a deterministic prediction by considering the augmented inputs $\psi_t:=\col{u_t,w_t}$.
\begin{prop}
Define a partition of $Z$ as
\begin{equation}
    Z:=\col{U,W,Y_p,Y_f}:=\col{\Psi,Y_p,Y_f},
\end{equation}
where $U\in\mathbb{R}^{n_u L\times M}$, $W\in\mathbb{R}^{n_w L\times M}$, $Y_p\in\mathbb{R}^{n_y L_0\times M}$, $Y_f\in\mathbb{R}^{n_y L'\times M}$, and $\Psi\in\mathbb{R}^{(n_u+n_w) L\times M}$. 
If $v_t=\mathbf{0}_{n_y}$ and $\text{rank}(Z)=(n_u+n_w) L+n_x$, we have
\begin{equation}
    \hat{\mathbf{y}}^t=Y_f\,g_\text{pinv}^t,\ g_\text{pinv}^t=\col{\Psi,Y_p}^\dagger\col{\mathbf{u}_\text{ini}^t,\hat{\mathbf{u}}^t,\mathbf{w}^t,\mathbf{y}_\text{ini}^t},
    \label{eqn:pinv}
\end{equation}
where $\hat{\mathbf{u}}^t:=(\hat{u}_k^t)_{k=0}^{L'-1}$ denotes the input sequence within the control horizon, $\mathbf{u}_\text{ini}^t:=(u_k)_{k=t-L_0}^{t-1}$, $\mathbf{y}_\text{ini}^t:=(y_k)_{k=t-L_0}^{t-1}$ denote the immediate past input and output sequences of length $L_0$, and $\mathbf{w}^t:=(w_k)_{k=t-L_0}^{t+L'-1}$ denotes the immediate past and future disturbance sequence of length $L$. The output sequence $\hat{\mathbf{y}}^t:=(\hat{y}_k^t)_{k=0}^{L'-1}$ provides a deterministic output prediction within the control horizon.
\label{prop:1}
\end{prop}
Indirect DDPC with deterministic predictor (\ref{eqn:pinv}) is known as subspace predictive control \citep{Favoreel_1999}.

\subsection{Stochastic Prediction}

In this work, two sources of uncertainties are considered. 1) Noise in output measurements. This induces noise in the output part of the signal matrix $Y_p$, $Y_f$, and the past output sequence $\mathbf{y}_\text{ini}^t$ with $\EE{\mathbf{y}_\text{ini}^t}=\bar{\mathbf{y}}_\text{ini}^t$, $\cov{\mathbf{y}_\text{ini}^t}=P_t$. 2) Uncertainties in the online disturbance sequence $\mathbf{w}^t$ with $\EE{\mathbf{w}^t}=\bar{\mathbf{w}}^t$, $\cov{\mathbf{w}^t}=\Sigma_w$. Statistics $\bar{\mathbf{w}}^t$ and $\Sigma_w$ can come from online measurements and predictions or prior knowledge.

Multiple algorithms have been developed to extend the deterministic prediction to the stochastic case. Such algorithms typically involve solving the following quadratic program:
\begin{equation}
\begin{split}
     g^t=\text{arg}\underset{g}{\text{min}}&\ \norm{Y_p g - \bar{\mathbf{y}}_\text{ini}^t}_S^2+\lambda\norm{g}_2^2\\
    \text{s.t.}&\quad  \Psi g=\col{    \mathbf{u}_\text{ini}^t,\hat{\mathbf{u}}^t,\bar{\mathbf{w}}^t},
\end{split}
\label{eqn:qp}
\end{equation}
where $\lambda$ and $S$ are design parameters. Different choices of $\lambda$ and $S$ have been proposed:
\begin{enumerate}[1)]
    \item \textit{Subspace predictor} \citep{fiedler2021relationship}: $S=\mathbb{I}_{n_y L_0}$, $\lambda\rightarrow 0^+$.
    \item \textit{Wasserstein distance minimization} \citep{lian2021adaptive}: $S=\mathbb{I}_{n_y L_0}$, $\lambda=n_y L_0\sigma^2$.
    \item \textit{Signal matrix model} \citep{yin2020maximum,pmlr-v144-yin21a}: $S=\mathbb{I}_{n_y L_0}$,
    \begin{equation*}
        \lambda=n_y \left(L\sigma^2+L'\sigma^2/\norm{g_\text{pinv}^t}_2^2\right).
    \end{equation*}
    \item \textit{Minimum mean-squared error} \citep{yin2021data}: $S=\bar{\Gamma}^\top \bar{\Gamma}$, $\lambda= n_yL'\sigma^2+\tr{S}\sigma^2$, where $\bar{\Gamma}$ is the last $n_y L_0$ columns of $Y_f\,\col{\Psi,Y_p}^\dagger$.
\end{enumerate}
See \cite{yin2021data} for a comparison between these choices. The quadratic program \eqref{eqn:qp} admits the following closed-form solution:
\begin{equation}
    g^t=
    \begin{bmatrix}
    R_1 & R_2 & R_3 & R_4
    \end{bmatrix}\col{\mathbf{u}_\text{ini}^t,\mathbf{u}^t,\bar{\mathbf{w}}^t,\bar{\mathbf{y}}_\text{ini}^t},
    \label{eqn:clg}
\end{equation}
where
\begin{align}
    \left[
    R_1\ R_2\ R_3\right]&:=F^{-1}\Psi^\top (\Psi F^{-1}\Psi^\top )^{-1},\label{eqn:r12}\\
    R_4&:=\left(F^{-1}-F^{-1}\Psi^\top (\Psi F^{-1}\Psi^\top )^{-1}\Psi F^{-1}\right)Y_p^\top S,\label{eqn:r3}
\end{align}
and $F:=\lambda\mathbb{I}_M+Y_p^\top S Y_p$.

The stochastic predictor can be constructed based on the solution $g^t$ with the following lemma.
\begin{lem}
The stochastic output sequence within the control horizon is given by 
\begin{equation}
\EE{\hat{\mathbf{y}}^t|\,g^t}=\bar{\mathbf{y}}^t,\ \cov{\hat{\mathbf{y}}^t|\,g^t}=\Sigma^t,    
\label{eqn:ydist}
\end{equation}
where
\begin{align}
    \bar{\mathbf{y}}^t&:=Y_f g^t - \Gamma(Y_p g^t - \bar{\mathbf{y}}_\text{ini}^t),\label{eqn:ybar}\\
    \Sigma^t&:=\Gamma P_t\Gamma^\top + \Gamma_w \Sigma_w\Gamma_w^\top + \norm{g^t}_2^2 T,\label{eqn:estat}\\
    T&:=\sigma^2\left(\Gamma\Gamma^\top+\mathbb{I}_{n_y L'}\right),\quad \Gamma_w:=(Y_f-\Gamma Y_p)R_3,\label{eqn:T}\\
    \Gamma&:=\col{CA^{L_0},\dots,CA^{L-1}}\col{C,\dots,CA^{L_0-1}}^\dagger
    \label{eqn:gam}
\end{align}
is the autonomous transformation matrix from $\mathbf{y}_\text{ini}^t$ to $\hat{\mathbf{y}}^t$.
\label{lm:1}
\end{lem}
\begin{pf}
This comes directly from the proof of Theorem~1 in \cite{yin2021data} by considering the augmented inputs $\psi_t$.\hfill$\square$
\end{pf}

Unfortunately, $\Gamma$ cannot be obtained exactly since $A$ and $C$ are unknown. However, this transformation matrix can also be estimated using a data-driven approach. Note that the true output prediction is given by $\hat{\mathbf{y}}^t_0=\Gamma \bar{\mathbf{y}}_\text{ini}^t$ if $\col{    \mathbf{u}_\text{ini}^t,\hat{\mathbf{u}}^t,\mathbf{w}^t}=\mathbf{0}$ and $P_t=\mathbf{0}$. Using the certainty equivalence principle, an estimate $\hat{\Gamma}_Z$ can be found by replacing $\hat{\mathbf{y}}^t_0$ with $\bar{\mathbf{y}}^t$. Then we have
\begin{equation}
    \bar{\mathbf{y}}^t=Y_f R_4 \bar{\mathbf{y}}_\text{ini}^t - \hat{\Gamma}_Z(Y_p R_4 \bar{\mathbf{y}}_\text{ini}^t - \bar{\mathbf{y}}_\text{ini}^t):=\hat{\Gamma}_Z\,\bar{\mathbf{y}}_\text{ini}^t,
\end{equation}
which leads to
\begin{equation}
    \hat{\Gamma}_Z=Y_f R_4\left(Y_p R_4\right)^{-1}.
    \label{eqn:gammahat}
\end{equation}
In what follows, it is assumed that $\Gamma=\hat{\Gamma}_Z$. This estimate is correct in the noise-free case and consistent under mild conditions as shown in the following propositions.
\begin{prop}
If $v_t=\mathbf{0}_{n_y}$, we have $\hat{\Gamma}_Z=\Gamma$.
\end{prop}
\begin{pf}
%This is a direct consequence of Lemma~2 in \cite{yin2021data}.
If $v_t=\mathbf{0}_{n_y}$, we have $\sigma^2=0$. All designs of $\lambda$ and $S$ are equivalent to the subspace predictor, under which case $R_4$ is the last $n_y L_0$ columns of $\col{\Psi,Y_p}^\dagger$ and thus $Y_p R_4=\mathbb{I}_{n_y L_0}$. Then Lemma~2 in \cite{yin2021data} directly leads to $\hat{\Gamma}_Z=\Gamma$.\hfill$\square$
\end{pf}
\begin{prop}
Let the singular values of $\col{\Psi,Y_p}$ be $\sigma_1,\dots,\sigma_{L_\sigma}$ in descending order, where $L_\sigma:=(n_u+n_w)L+n_yL_0$. Then as $M\rightarrow\infty$, $\hat{\Gamma}_Z\rightarrow\Gamma$ w.p. 1, if $\sigma_{L_\sigma}\rightarrow\infty$.
\end{prop}
\begin{pf}
Let $\col{\Psi,Y_p}:=\Omega SV^\top$ be the singular value decomposition, where $\Omega,S\in\mathbb{R}^{L_\sigma\times L_\sigma}$ and $V\in\mathbb{R}^{M\times L_\sigma}$. Then, $g_\text{pinv}^t=VS^{-1}\Omega^\top \omega^t$, where $\omega^t:=\col{\mathbf{u}_\text{ini}^t,\mathbf{u}^t,\bar{\mathbf{w}}^t,\bar{\mathbf{y}}_\text{ini}^t}$, and $\norm{g_\text{pinv}^t}_2^2\leq \norm{V}_2^2\norm{S^{-1}}_2^2\norm{\Omega}_2^2\norm{\omega^t}_2^2=\norm{\omega^t}_2^2 {\big /} \sigma_{L_\sigma}^2$.
Note that $g_\text{pinv}^t$ is also the least-norm solution to the linear system $\col{\Psi,Y_p}g=\omega^t$, so we have $\norm{g^t}_2^2\leq\norm{g_\text{pinv}^t}_2^2$. Therefore, if $\sigma_{L_\sigma}\rightarrow\infty$, $\norm{g^t}_2^2\rightarrow 0$ and $\Sigma^t\rightarrow \mathbf{0}$ since $P_t=\mathbf{0}$. This directly leads to the convergence of $\hat{\Gamma}_Z$ to $\Gamma$.
\vspace{-0.5em}
\ \hfill$\square$
\end{pf}
\begin{rem}
The singular value condition $\sigma_{L_\sigma}\rightarrow\infty$ requires that the columns of $\col{\Psi,Y_p}$ activate all directions persistently as $M\rightarrow\infty$. This is satisfied for, for example, independent random or repeated full-rank inputs and disturbances.
\end{rem}

\section{Stochastic Indirect Data-Driven Predictive Control}
\label{sec:main}

Based on the stochastic predictor (\ref{eqn:ydist}), the stochastic indirect DDPC algorithm can be proposed. In the following subsections, the stochastic control cost is first formulated as a quadratic objective. Then, the prediction accuracy is improved by filtering the output initial condition estimates with a Kalman filter. Finally, the satisfaction of chance constraints is guaranteed by formulating tightened SOC constraints.

\subsection{Stochastic Control Cost}

The stochastic control cost $J_t$ is formulated as a quadratic function in the following lemma.
\begin{lem}
The expected control cost is given by
\begin{equation}
    J_t = \norm{\hat{\mathbf{u}}^t}^2_{\bar{R}} + \norm{\bar{\mathbf{y}}^t-\mathbf{r}^t}^2_{\bar{Q}} + \tr{\bar{Q}T}\norm{g^t}_2^2 +\text{constant},
\label{eqn:jt}
\end{equation}
where $\bar{R} := \mathbb{I}_{L'}\otimes R$, $\bar{Q} := \mathbb{I}_{L'}\otimes Q$, $\mathbf{r}^t:=(r_{t+k})_{k=0}^{L'-1}$ and $T:=\sigma^2\left(\Gamma\Gamma^\top+\mathbb{I}_{n_y L'}\right)$. The cost is quadratic with respect to the optimization variable $\hat{\mathbf{u}}^t$.
\end{lem}
\begin{pf}
The expected output cost is calculated as:
\begin{align*}
    &\EE{\norm{\hat{\mathbf{y}}^t-\mathbf{r}^t}_{\bar{Q}}^2}=\EE{\left(\bar{\mathbf{y}}^t+\mathbf{e}^t-\mathbf{r}^t\right)^\top \bar{Q}\left(\bar{\mathbf{y}}^t+\mathbf{e}^t-\mathbf{r}^t\right)}\\=&\norm{\bar{\mathbf{y}}^t-\mathbf{r}^t}_{\bar{Q}}^2 + \mathbb{E}\left[\left(\mathbf{e}^t\right)^\top {\bar{Q}}\,\mathbf{e}^t\right]=\norm{\bar{\mathbf{y}}^t-\mathbf{r}^t}_{\bar{Q}}^2 + \mathrm{tr}\left({\bar{Q}}\Sigma^t\right)\\
    =&\norm{\bar{\mathbf{y}}^t-\mathbf{r}^t}_{\bar{Q}}^2 + \mathrm{tr}\left({\bar{Q}}T\right)\norm{g^t}_2^2+\mathrm{const},
\end{align*}
where $\mathbf{e}^t$: $\EE{\mathbf{e}^t}=\mathbf{0}$, $\cov{\mathbf{e}^t}=\Sigma^t$ is the prediction error. The second to last equality is due to the cyclic property of the trace function. This cost is quadratic with respect to $\hat{\mathbf{u}}^t$ since both $g^t$ and $\bar{\mathbf{y}}^t$ are linear with respect to $\hat{\mathbf{u}}^t$.\hfill$\square$
\end{pf}
The stochastic control cost adds a $\norm{g^t}_2^2$-regularization term to the nominal cost. Such regularization is required in direct DDPC \citep{Coulson_2019} for well-definedness and is proposed to enhance robustness in indirect DDPC \citep{pmlr-v144-yin21a}. However, it was unclear how to tune the weighting factor for the regularizer other than trial and error. By considering the regularizer as the uncertainty term in the expected output cost, the weighting factor can be reliably selected as $\mathrm{tr}\left({\bar{Q}}T\right)$, which depends on the output cost matrix and the noise level.

\subsection{Initial Condition Estimation}

In model-based output-feedback MPC, an estimator has to be designed to estimate the initial state of the predictor, which is not measurable. This is not required in DDPC since the output initial condition $\bar{\mathbf{y}}_\text{ini}^t$ can be directly measured. In fact, in most existing DDPC implementations with stochastic data, the output initial condition $\bar{\mathbf{y}}_\text{ini}^t$ comes from measurements as in the deterministic case, i.e., $\bar{\mathbf{y}}_\text{ini}^t:=(y_k)_{k=t-L_0}^{t-1}$. Thus, the associated covariance $P_t=\sigma^2\mathbb{I}$ is constant in \eqref{eqn:estat}. This source of uncertainty can be alleviated by choosing a larger $L_0$ \citep{pmlr-v144-yin21a}. On the other hand, in the presence of stochastic uncertainties, a properly designed estimator can estimate the initial state with a diminishing covariance that is much smaller than the noise level in the measurements.

Therefore, although not required, it can be beneficial to improve the output initial condition measurements based on output predictions at previous time steps by designing an estimator, especially in cases where the online measurement error is large. In this subsection, a Kalman filter is designed as the estimator. In particular, we replace $y_k$ with its Kalman-filtered counterpart for the output initial condition. This reduces the prediction errors by shrinking $P_t$ as time progresses.

In detail, the same predictor (\ref{eqn:ydist}) for predictive control design at time $(t-1)$ is used to filter the output at time $t$ and update $\bar{\mathbf{y}}_\text{ini}^t$. The predictor can be considered as a non-minimal state-space ``model'' with ``state''
\begin{equation}
    \bar{x}_t:=\text{col}\left(u_{t-L_0},\dots,u_{t-1},y_{t-L_0}^0,\dots,y_{t-1}^0\right).
\end{equation}
Let $\bar{y}_i^t$ and $e_i^t$ denote the $(i+1)$-th block element of $\bar{\mathbf{y}}^t$ and $\mathbf{e}^t$, respectively, and $\Sigma^t_i$ be the covariance of $e_i^t$, i.e., the $(i+1)$-th $n_y\times n_y$ block on the diagonal of $\Sigma^t$. The data-driven ``model'' is then given by
\begin{equation}
\begin{cases}
\bar{x}_{t+1}&=\ \underbrace{\begin{bmatrix}\Lambda^{n_u}&\mathbf{0}\\\mathbf{0}&\Lambda^{n_y}\end{bmatrix}}_{\bar{\Lambda}} \bar{x}_t+\begin{bmatrix}\mathbf{0}\\\hat{u}_0^t\\\mathbf{0}\\\bar{y}_0^t+e_0^t\end{bmatrix},\\
\hfil \zeta_{t+1}&=\ \begin{bmatrix}\mathbf{0}&\mathbb{I}_{n_y}\end{bmatrix}\bar{x}_{t+1}+v_t=y_t^0+v_t=y_t,
\label{eq:sys2}
\end{cases}
\end{equation}
where $\Lambda^k$ denotes the $k$-step upper shift matrix with ones on the $k$-th superdiagonal. The covariances of the ``process noise'' $e_0^t$ and the measurement noise $v_t$ are $\Sigma^t_0$ and $\sigma^2\mathbb{I}_{n_y}$, respectively. Then, a Kalman filter for \eqref{eq:sys2} can be designed to estimate the initial condition $\bar{x}_t$. Let the state estimate and the output part of the state error covariance be $\bar{x}_{t,t}$ and $P_{t,t}$, respectively. Then, the initial conditions for the DDPC problem can be set as $\col{\mathbf{u}_\text{ini}^t,\bar{\mathbf{y}}_\text{ini}^t}:=\bar{x}_{t,t}$ and $P_t := P_{t,t}$. The Kalman filtering algorithm is summarized in Algorithm~\ref{al:1}.

\begin{algorithm}[htb]
	\caption{Kalman filter in stochastic indirect DDPC}
	\begin{algorithmic}[1]
    \State\textbf{Initialization:}
    \begin{align}
    \bar{x}_{0,0}&:=\text{col}\left(u_{-{L_0}},\dots,u_{-1},y_{-{L_0}},\dots,y_{-1}\right),\\ P_{0,0}&:=\mathbb{I}_{n_y {L_0}}.
    \end{align}
    \State\textbf{Prediction:}
    \begin{align}
        \bar{x}_{t,t+1}&:=\bar{\Lambda}\bar{x}_{t,t}+\text{col}\left(\mathbf{0},\hat{u}_0^t,\mathbf{0},\bar{y}_0^t\right),\\
        P_{t,t+1}&:=\Lambda^{n_y}P_{t,t}\left(\Lambda^{n_y}\right)^\top +\Sigma^t_0.
    \end{align}
    \State\textbf{Update:}
    \begin{align}
        K_{t+1}&:=\Sigma^t_0\left(\Sigma^t_0+\sigma^2\mathbb{I}_{n_y}\right)^{-1},\\
        \bar{x}_{t+1,t+1}&:=\bar{x}_{t,t+1}+\text{col}\left(\mathbf{0},K_{t+1}\left(y_t-\bar{y}_0^t\right)\right),\\
        P_{t+1,t+1}&:=\left(\mathbb{I}_{n_y}-K_{t+1}\right)P_{t,t+1}.
    \end{align}
	\end{algorithmic}
	\label{al:1}
\end{algorithm}

\begin{rem}
Only one-step ahead prediction is required to run the Kalman filter. Here, it is obtained by truncating the same $L'$-step ahead predictor used in predictive control for simplicity. One can also similarly construct a one-step ahead data-driven predictor with $L'=1$, specifically for the Kalman filter.
\end{rem}

\begin{rem}
A similar idea was proposed in \cite{Alpago_2020} for a direct DDPC algorithm. However, no approach is provided to quantify the covariance of the prediction error required in the Kalman filter as there is no well-defined predictor in direct DDPC.
\end{rem}

\subsection{Chance Constraint Satisfaction}

As mentioned in Section~\ref{sec:2}, the output constraints $y_t\in\mathcal{Y}_t$ cannot be guaranteed robustly under unbounded noise. Instead, high-probability chance constraints are considered, either element-wise as
\begin{equation}
    \text{Pr}({h_i^{t+k}}^\top \hat{y}_k^t\leq q_i^{t+k})\geq p,\ \forall\,i=1,\dots,n_c, k=0,\dots,L'-1,
    \label{eqn:cc1}
\end{equation}
or set-wise as
\begin{equation}
    \text{Pr}\left(\hat{y}_k^t\in \mathcal{Y}_{t+k}\right)\geq p,\ \forall\,k=0,\dots,L'-1,
    \label{eqn:cc2}
\end{equation}
where $p$ is the targeted probability. These chance constraints are typically guaranteed by tightening the nominal constraints to account for prediction uncertainties. However, unlike standard model-based predictors with additive uncertainties, the prediction error covariance of the data-driven predictor (\ref{eqn:ydist}) depends on the particular inputs and initial conditions via $g^t$. So the amount of constraint tightening cannot be calculated offline. Define the augmented linear constraints by $\bar{\mathcal{Y}}_t=\setdef{\mathbf{y}}{\bar{H}^t\mathbf{y}\leq \bar{q}^t}$, where
\begin{align*}
    \bar{H}^t&:=\left[\bar{h}^t_1\ \dots\ \bar{h}^t_{L'n_c}\right]^\top:=\text{blkdiag}\left(H^t,\dots,H^{t+L'-1}\right),\\
    \bar{q}^t&:=\col{\bar{q}^t_1,\dots,\bar{q}^t_{L'n_c}}:=\col{q^t,\dots,q^{t+L'-1}}.
\end{align*}
The following lemma guarantees chance constraint satisfaction by constraint tightening.
\begin{lem}
The constraint
\begin{equation}
    \bar{q}^t - \bar{H}^t\bar{\mathbf{y}}^t \geq\mu\sqrt{\text{diag}\left(\bar{H}^t\Sigma^t{\bar{H}^{t}}^\top\right)}
    \label{eqn:soccon}
\end{equation}
guarantees the satisfaction of the chance constraints \eqref{eqn:cc1} if $\mu \geq \sqrt{\tfrac{1}{1-p}-1}$ and \eqref{eqn:cc2} if $\mu \geq \sqrt{\tfrac{n_y}{1-p}}$.
\label{lm:3}
\end{lem}
\begin{pf}
Applying the one-sided Chebyshev's inequality, we have
\begin{equation}
    \text{Pr}\left(\bar{h}_i^t \hat{\mathbf{y}}^t-\bar{h}_i^t \bar{\mathbf{y}}^t\leq \sqrt{\tfrac{1}{1-p}-1}\cdot\text{std}\left(\bar{h}_i^t \hat{\mathbf{y}}^t\right)\right)\geq p,\ \forall\,i,
    \label{eqn:cheb}
\end{equation}
where $\text{std}\left(\bar{h}_i^t \hat{\mathbf{y}}^t\right)=\sqrt{\bar{h}_i^t\Sigma^t{\bar{h}_i^t}^\top}$. From \eqref{eqn:soccon}, we have
\begin{equation}
    \bar{q}^t_i - \bar{h}^t_i\bar{\mathbf{y}}^t \geq\mu\sqrt{\bar{h}_i^t\Sigma^t{\bar{h}_i^t}^\top}.
    \label{eqn:soccon2}
\end{equation}
Equations \eqref{eqn:cheb} and \eqref{eqn:soccon2} lead to \eqref{eqn:cc1} for $\mu \geq \sqrt{\tfrac{1}{1-p}-1}$.

From the multi-dimensional Chebyshev's inequality, the ellipsoidal set
\begin{equation}
    \mathcal{E}_t:=\setdef{\mathbf{e}}{\mathbf{e}^\top \left(\Sigma^t\right)^{-1}\mathbf{e}\leq \frac{n_y}{1-p}}
\end{equation}
is a confidence region of prediction error $\mathbf{e}^t$ with at least probability $p$. Then, the chance constraint is satisfied if
\begin{equation}
    \bar{\mathbf{y}}^t\in \bar{\mathcal{Y}}_t\ominus \mathcal{E}_t:=\setdef{\mathbf{y}}{\mathbf{y}+\mathbf{e}\in\bar{\mathcal{Y}}_t,\forall\,\mathbf{e}\in\mathcal{E}_t},
\end{equation}
where $\ominus$ denotes the Pontryagin difference. For polytope $\bar{\mathcal{Y}}_t$ and ellipsoid $\mathcal{E}_t$, we have
\begin{equation}
    \bar{\mathcal{Y}}_t\ominus\mathcal{E}_t=\setdef{\mathbf{y}}{\bar{h}_i^\top\mathbf{y}\leq \bar{q}_i-\eta_{\mathcal{E}_t}\left(\bar{h}_i\right), i=1,\dots,L'n_c},
\end{equation}
where $\eta_{\mathcal{E}_t}\left(\bar{h}_i\right):=\sqrt{\tfrac{n_y}{1-p}\bar{h}_i^\top\Sigma^t\bar{h}_i}$ is the support function of $\mathcal{E}_t$. Aggregating the constraints for all $i$ leads to \eqref{eqn:cc1} for $\mu \geq \sqrt{\tfrac{n_y}{1-p}}$.\hfill$\square$
\end{pf}
\begin{rem}
    Let $F_{\chi^2_d}(\cdot)$ and $F_{\mathcal{N}}(\cdot)$ be the cumulative distribution function of the $\chi^2$-distribution with $d$ degrees of freedom and the unit Gaussian distribution, respectively. The lemma can be tightened if Gaussian uncertainties are considered, i.e., both $v_t$ and $\mathbf{w}^t$ are Gaussian, by choosing $F_{\mathcal{N}}(\mu)\geq p$ for \eqref{eqn:cc1} and $F_{\chi^2_{n_y}}(\mu^2)\geq p$ for \eqref{eqn:cc2}, respectively. The proof is very similar to that of Lemma~\ref{lm:3}.
\end{rem}
Unfortunately, the tightened constraint \eqref{eqn:soccon} is not convex. The following corollary provides a convex surrogate of \eqref{eqn:soccon}.
\begin{cor}
    The SOC constraint
    \begin{equation}
    \bar{q}^t - \bar{H}^t\bar{\mathbf{y}}^t \geq\mu\left(\mathbf{c}_1+\mathbf{c}_2\norm{g^t}_2\right),
    \label{eqn:soccon3}
    \end{equation}
    where
    \begin{align}
    \mathbf{c}_1&:=\sqrt{\text{diag}\left(\bar{H}^t\left(\Gamma\Sigma_y\Gamma^\top + \Gamma_w \Sigma_w\Gamma_w^\top\right){\bar{H}^{t}}^\top\right)},\\
    \mathbf{c}_2&:=\sqrt{\text{diag}\left(\bar{H}^t T {\bar{H}^{t}}^\top\right)},
    \end{align}
    guarantees the satisfaction of \eqref{eqn:soccon}.
\end{cor}
\begin{pf}
Since $\sqrt{\sum_i a_i}\leq\sum_i\sqrt{a_i}$, we have
\begin{equation*}
    \mathbf{c}_1+\mathbf{c}_2\norm{g^t}_2 \geq \sqrt{\mathbf{c}_1^2 + \mathbf{c}_2^2\norm{g^t}_2^2}=\sqrt{\text{diag}\left(\bar{H}^t\Sigma^t{\bar{H}^{t}}^\top\right)}.\ \ \square
\end{equation*}
\end{pf}

The proposed stochastic indirect DDPC algorithm is summarized in Algorithm~\ref{al:2}.
\begin{algorithm}[tb]
	\caption{Stochastic indirect DDPC}
	\begin{algorithmic}[1]
	\State Select a data-driven predictor and calculate predictor parameters from \eqref{eqn:r12}, \eqref{eqn:r3}, \eqref{eqn:T}, and \eqref{eqn:gammahat}.
	\State Initialize the Kalman filter from Algorithm~\ref{al:1}.
	\For{$t\leftarrow 0,1,\dots$}
	\State $\col{\mathbf{u}_\text{ini}^t,\mathbf{y}_\text{ini}^t}\leftarrow \bar{x}_{t,t}$, $P_t \leftarrow P_{t,t}$
	\State $\hat{\mathbf{u}}^t\leftarrow\text{arg}\underset{\hat{\mathbf{u}}^t}{\text{min}}\ \ \eqref{eqn:jt}\quad\text{s.t. }\ \eqref{eqn:clg},\eqref{eqn:ybar},\eqref{eqn:ucon},\eqref{eqn:soccon3}$.
	\State Apply $u_t=\hat{u}_0^t$ to the system and measure $y_t$.
	\State Run the Kalman filter from Algorithm~\ref{al:1}.
	\EndFor
	\end{algorithmic}
	\label{al:2}
\end{algorithm}

\section{Numerical Example}
In this section, we compare the performance of nominal DDPC (\textit{N-DDPC}), DDPC with initial condition estimation in Algorithm~\ref{al:1} (\textit{KF-DDPC}), and stochastic DDPC in Algorithm~\ref{al:2} (\textit{S-DDPC}). Consider the following fourth-order dynamics:
\begin{equation*}
\left[
\begin{array}{c|c|c}
A & B & E \\
\hline
C & 0 & 0
\end{array}
\right]=
\left[
\begin{array}{cccc|c|c}
 0.36 &  0.64 & 0.07 &  0.02 & 0.29 & 0.03 \\
 0.42 &  0.58 & 0.02 &  0.07 & 0.03 & 0.20 \\
-9.34 &  9.34 & 0.23 &  0.58 & 4.90 & 1.07 \\
 5.88 & -5.88 & 0.39 & -0.39 & 1.07 & 3.48 \\
\hline
1 & 0 & 0 & 0 & 0 & 0
\end{array}
\right].
\end{equation*}
The following parameters are used in the example: $L_0=4$, $L'=10$, $Q=20$, $R=1$, $\sigma^2=0.01$, $p=0.95$. An offline trajectory of length 500 is collected with unit Gaussian inputs and the signal matrix is constructed with a Hankel structure, which leads to $M=487$. The statistics of the disturbance are given by $\bar{\mathbf{w}}^t=\mathbf{0}$ and $\Sigma_w = 0.001\cdot\mathbb{I}$. The elementwise chance constraints \eqref{eqn:cc1} are used. The same online noise and disturbance sequences are used to compare the three algorithms. The minimum mean-squared error predictor in \cite{yin2021data} is employed as the predictor. No input constraint is considered in this example, i.e., $\mathcal{U}_t=\mathbb{R}$. Upper and lower output bounds are specified as the output constraints.

\begin{figure}[tb]
    \centering
    \includegraphics[width=0.95\columnwidth]{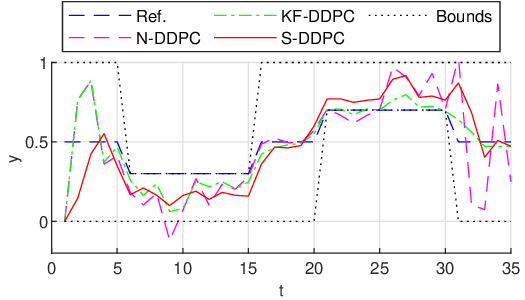}
    \vspace{-1em}
    \caption{Closed-loop trajectories of DDPC algorithms.}
    \label{fig:1}
\end{figure}

\begin{figure}[tb]
    \centering
    \includegraphics[width=0.95\columnwidth]{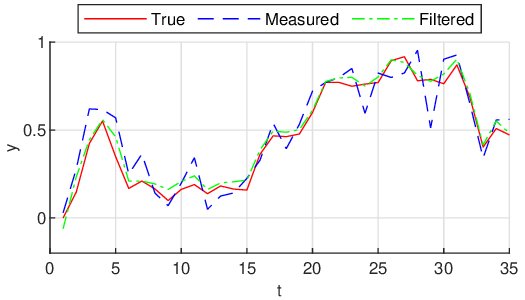}
    \vspace{-1em}
    \caption{Comparison of filtered and measured output trajectories.}
    \label{fig:2}
\end{figure}

The closed-loop trajectories of the algorithms are presented in Figure~\ref{fig:1}, alongside the reference trajectory and the output bounds. As observed in Figure~\ref{fig:1}, \textit{KF-DDPC} outperforms \textit{N-DDPC} by introducing the initial condition estimator, although constraint violations are still evident. \textit{S-DDPC} further enhances \textit{KF-DDPC}, particularly in terms of constraint satisfaction. To underscore the effectiveness of the Kalman filter, Figure~\ref{fig:2} showcases the comparison between the filtered output initial conditions and the measured ones for \textit{S-DDPC}. The filtered trajectory is notably closer to the true trajectory compared to the measured trajectory.

\begin{figure}[tb]
    \centering
    \includegraphics[width=0.475\columnwidth]{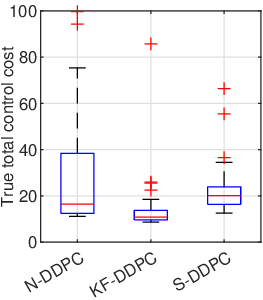}
    \includegraphics[width=0.475\columnwidth]{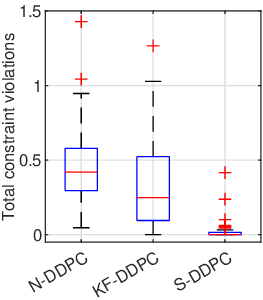}
    \vspace{-0.25em}
    {\footnotesize(a)}\qquad\qquad\qquad\qquad\qquad\quad\ {\footnotesize(b)}
    \vspace{-0.25em}
    \caption{Boxplots of (a) the true total control cost and (b) the total amount of constraint violations.}
    \label{fig:3}
\end{figure}

The performance is further evaluated quantitatively by 50 Monte Carlo simulations with different noise and disturbance realizations. Figure~\ref{fig:3} shows the boxplots of the true total control cost and the total amount of constraint violations, calculated as $\sum_t \max\left(H^t y_t - q^t,0\right)$. The results validate our observations from Figure~\ref{fig:1} that our proposed algorithm \textit{S-DDPC} performs much better than the nominal algorithm with almost no constraint violation.

\section{Conclusion}
This work discusses several modifications in stochastic data-driven predictive control (DDPC) algorithms. They provide a tuning-free regularizer design in the control cost, improved initial condition estimation, and reliable constraint satisfaction. These are achieved by evaluating the expected cost, designing a Kalman filter, and formulating convex constraint tightening terms, respectively. These modifications pave the way for providing theoretical guarantees for DDPC algorithms under general unbounded stochasticity.

\bibliography{refs}             % bib file to produce the bibliography

\end{document}